\documentclass[english,aps,pra,reprint,showpacs]{revtex4-1}   

\usepackage[T1]{fontenc}	

\usepackage[latin9]{inputenc}	

\usepackage{geometry}		

\usepackage{graphicx}

\usepackage[above,below]{placeins}	

\usepackage{times}

\usepackage{color}
\newcommand{\ecs}[1]{\colorbox{yellow}{ECS: #1}}

\begin{document}

\title{``Pulling out'' as a procedural resource when solving partial differential equations}

\author{Bahar Modir}


\author{Eleanor C. Sayre}

\affiliation{Department of Physics, Kansas State University, 116 Cardwell Hall, Manhattan, Kansas, USA, 66506-2601}


\begin{abstract}

We investigate how students solve partial differential equations and partial derivatives in the context of quantum mechanics. We use the resources framework to investigate students' discussion in a group problem-solving environment to investigate the fine-grain elements of their problem solving. We analyze an example of students' use of separation of variables to solve a partial differential equation for a free particle problem. We identify a mathematical action called ``pulling out'' as a procedural resource to help students with separating the time part from the space part of the wave function in the course of solving the time-dependent Schr{\"o}dinger equation. We discuss how students use ``pulling out'' as a procedural step in solving partial differential equations and sense-making.

\end{abstract}


\maketitle

\section{Introduction}

We are interested in how students use mathematics to solve partial differential equations (PDEs) and perform partial derivatives (PDs) across upper-division physics courses. PDEs are endemic in upper-division physics theory courses, from the Maxwell relations in thermodynamics, to Maxwell's equations in electromagnetism,  and the Schr{\"o}dinger equation in quantum mechanics.  Previous research on student understanding of PDEs has focused primarily on thermodynamics\cite{Roundy2014,Kustusch2014}.  

To understand how students solve PDEs and perform PDs, we turn to the resources framework.  Resources are small, reusable elements of student reasoning with internal structure\cite{Hammer2000,Sayre2008}. Students use procedural resources\cite{Black2007,Wittmann2015} to perform actions such as separating variables to solve a PDE.  In this paper, we present a specific procedural resource -- \textit{pulling out} -- which students use to solve PDEs and perform PDs.  Though the work we present here is in quantum mechanics, we believe \textit{pulling out} is broadly applicable across coursework in physics.

Wittmann et al.,~\cite{Wittmann2015} investigated procedural resources such as \textit{separation}, finding that their internal structure includes several resources: \textit{grouping, division} or \textit{multiplication, moving} and \textit{subtraction}, explorable by students. However, the level of structure varies among students with different level of expertise. Students can use  \textit{separation} as a single resource, or they can activate additional resources to successfully separate variables at both sides of an equation. They reported that intermediate mechanics students struggle to group terms properly on each side of the equation.

In this paper, we present two examples from the same group of three students enrolled in an upper-level quantum mechanics classroom.  The students work together during class to obtain the Time Independent Schr{\"o}dinger Equation (TISE) for the space part,  starting with the Time Dependent Schr{\"o}dinger Equation (TDSE).  We analyze two in-class problem-solving sessions, one near the beginning of the course and one two months later.  We closely examine their discussions to show the existence of the \textit{pulling out} resource, and to build resource graphs using and unpacking the \textit{separation of variables} resource in this context.

\section{Context}

We collected video data from one semester of a senior-level undergraduate quantum mechanics course. The class was a mixture of traditional lecture and spontaneous bursts of in-class problem solving where students worked in groups of 2-3 to solve the questions on the whiteboards. We collected video data of students' whiteboards' work for three different groups through out the semester. The instructor controlled the length of each problem solving interlude, generally 2-5 minutes. The textbook used was Griffiths' Introduction to Quantum Mechanics~\cite{griffiths2005}. To determine students' use of mathematics is physics, we closely examined students' discourse, gesture and whiteboard writing as they worked on a problem to provide evidence of their use of procedural resources in their solutions. 

 Since the problem the students are solving is chosen from the context of TDSE ($H\Psi(\textbf{r} , t)=i\hbar\frac{\partial\Psi(\textbf{r} , t)}{\partial t}$), so we review the physics and mathematics aspects briefly for the reader. The TDSE is a partial differential equation that can describe the time evolution of any physical system with different potentials and boundary conditions. The Hamiltonian can be written in terms of the  Laplacian and the potential. By choosing the natural coordinate of a physical system, one can expand the Laplacian. The easiest problem to consider is the free particle in one dimension of space $x$. In this case, the Hamiltonian has only the kinetic energy term, and the partial differential equation depends on two variables of $x$ and $t$. One way to solve a partial differential equation is to break it into a series of independent one variable equations and solve each of those separated equations. This method is called separation of variables (SOV). The SOV condition is to assume that the total wave function is a product of independent one variable wave functions. By substituting the total wave function in to the TDSE, the partial derivatives can turn into ordinary derivatives and after a couple of mathematical procedures the equation becomes separable. The separated spatial part is in the form of $H\psi(\textbf{r})=E\psi(\textbf{r})$.

\section{First interaction}

One student,``Alex,'' is very comfortable solving TDSE via SOV: he treats some of the procedures as trivial, and he does not explicitly mention those steps. Alex uses SOV as an entity which is very compressed. However, he is mindful that they are necessary steps and brings them into play, when other students ask him questions to explain some of the skipped steps. In this paper we will treat the interaction between Alex and Eric as a case of ``unpacking'' resources. As Eric seeks for more elaboration, Alex gradually unpacks his unit of SOV (see Fig.\ \ref{fig1}). This gives us an opportunity to identify the resources that students bring into play.

\subsection{Grouping and dividing}

On his first pass through the problem, Eric starts to write the TDSE in the form of an differential equation acting on $\psi$. As Eric writes the first line, Alex changes the small $\psi$ symbol to the $\Psi(x,t)$, and also puts a negative sign on the right side of Eric's equation. Eric finishes writing the TDSE by setting the potential term equal to zero. They both will not notice a missing factor of $\frac{-1}{2m}$ on the right side of the equation, which will not affect the correctness of their final solution. He asserts that the space solution will be sine waves, saying dismissively ``Oh well this is, like just a sine waves, something like that.''

At this point, he does not go through all the steps of solving the ordinary differential equation.  Alex interrupts Eric's solution train midway to use a more formal separation of variables procedure:


%
%
%

\begin{description} 

\vspace{-.2cm}

\item[Alex] Eh\dots We have separating, so we can have them as separable $\psi$, Capital $\Psi$ is $\psi(x)$ times function of time (writing $\Psi(x,t)$=$\psi(x){f(t)}$)

\vspace{-.3cm}

\item[Eric] Ok\dots oh I see, I see\dots oh yea so then these two each must equal a constant because\dots right?

\vspace{-.3cm}

\item[Alex] This is only a function of time (pointing to the left side of the separated TDSE)

\vspace{-.3cm}

\item[Alex and Eric] (Eric becomes in unison with Alex) And that's only a function of ${x}$ (both pointing to the right side of the TDSE)

\item[Alex] Well it will be when we divide by capital $\Psi$, so we get, so we get that $\frac{f'}{f}$ =  $-i\hbar\frac{\psi''}{\psi}$, once you divide through.

\vspace{-.2cm}

\end{description}

\begin{figure}

\includegraphics[width=0.8\linewidth]{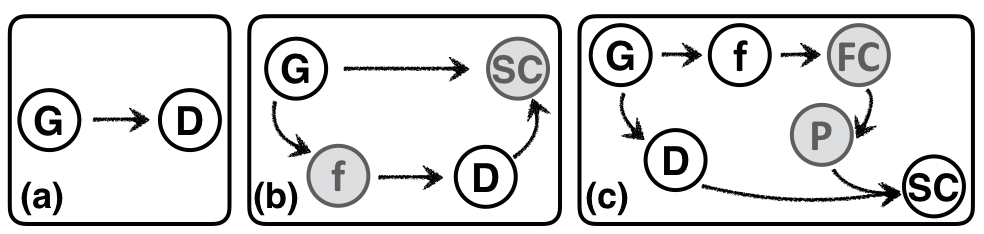}

\caption{Phases of unpacking separation of variables.  a: initial unpacking. b: intermediate unpacking. c: final unpacking. Key in Table \ref{table1}
\label{fig1}}

\vspace{-.6cm}
\end{figure}

Eric readily goes along with and elaborates on Alex's bid to use separation of variables more formally.  Together, they group space functions on the right hand side of the equation and time functions on the right.  In his last statement, Alex references that if $\Psi(x,t)$ is the product of two $\psi(x){f(t)}$, the PDE is separable by dividing both sides by $\Psi$.  At this point in their joint explanation, they are using two procedural resources to build the separation of variable resource: \textit{grouping} terms by variable, and \textit{dividing} both sides by the original function $\Psi$ (figure \ref{fig1}a). Neither student has talked about a separation constant or what its role might be in the solution, and they have not explicitly combined their TDSE with the separable form for $\Psi$.

\begin{table}[htp]
\caption{Resources in unpacked \textit{separation of variables}\label{table1}}
\begin{center}
\begin{tabular}{|c|l|c|}
\hline
Abbr&Name&Type\\
\hline
G&\textit{grouping}&procedural\\
D&\textit{dividing}&procedural\\
f&\textit{function}&conceptual\\
SC&\textit{separation constant}&conceptual\\
FC&\textit{functions-as-constants}&conceptual\\
P&\textit{pulling out}&procedural\\
\hline
\end{tabular}
\end{center}
\end{table}%

\subsection{Functions and separation constants}

As the students continue with their solution, Eric asks questions to elaborate on and further unpack the \textit{separation of variables} procedural resource.  

After seven seconds of silence, Eric points to the $\Psi(x,t)$ in the TDSE and says ``oh  $\psi$ is double primed''. Although he does not mention $\Psi(x,t)$, it seems that Eric still cannot distinguish between $\Psi(x,t)$ and $\psi(x)$. 

Alex explains further: he adds two more pieces to his previous explanation \textit{separation of variables} (see Fig.\ \ref{fig2}b). Alex briefly explains that the $\Psi(x,t)$ will be divided through both sides of the equation, and since each side is a product of a function multiplied by the derivative of another function, the division results into two separated equations equalling ``some constant''. Even before highlighting the functions and the derivations in terms of $\psi(x)$ and ${f(t)}$ functions, Alex feels that everything was ready to be completely separated by just one action (division). 

However, Alex again points to the TDSE to explain, how after substitution of the product of two functions, what each side of the equation looks like, and how division leads each side of the equation becoming a group of a single variable.






\begin{description}

\vspace{-.2cm}

\item[Eric] Oh $\psi$ is double primed\dots 

\vspace{-.3cm}

\item[Alex] Because this is ${\psi''}$ times ${f}$ and this is ${f''}$ times $\psi$ (pointing to each side of the TDSE) and then [unintelligible] divide through by capital $\Psi$ (pointing to the whole TDSE with gesture of hands with motion)\dots and then you get this (pointing to the separated equation) 

\vspace{-.3cm}

\item[Eric] Okey okey

\vspace{-.2cm}

\end{description}

We interpret Alex's explicitly unpacking of the functional relationship between $\Psi$, $\psi$, and $f$ as activating the \textit{function} resource and inserting it into the \textit{separation of variables} procedure between \textit{grouping} and \textit{dividing}.

Alex continues immediately with explaining that each side of the PDE ``equals $k$\dots some constant''.  This is the first mention of a separation constant in the interaction.  We interpret this as the activation and insertion of another resource into the graph for the \textit{separation of variables} procedure (figure \ref{fig1}b). Alex does not further explain the procedures; what he has held constant in each derivative, or with respect to what variable the derivation is taken. 


%
%

In this part of the conversation Alex unpacked \textit{separation of variables} more, inserting two more resources: \textit{function} in the form of $\psi$ and $f$, and \textit{separation constant} in the form of $k$. These two resources are conceptual, not procedural, and serve to explain how \textit{grouping} and \textit{dividing} are connected.

\subsection{Pulling out functions as constants}

Eric is still a bit confused about how dividing both sides of the TDSE by $\Psi(x,t)$ results in the final separated solution. He again returns to the TDSE and points to the $\Psi(x,t)$ on one side of the TDSE, where Alex had pointed out to that location earlier to write down the $\Psi(x,t)$ as a product of two functions. He asks an important question, pointing at the right side of the unseparated TDSE: ``How did you, like, get to that?''  He's asking Alex to explain a missing procedure whereby a partial time derivative of $\Psi(x,t)$ becomes a time derivative of only $f(t)$.  

\begin{figure}
\includegraphics[width=0.8\linewidth]{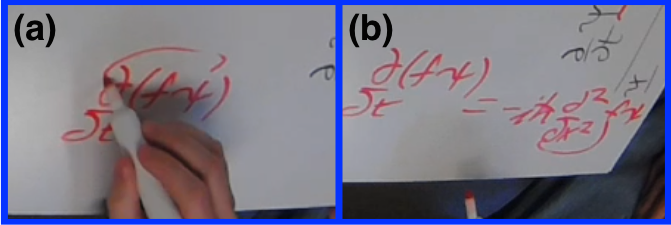}
\caption{$(a)$ Pulling out the space part out of the time part. $(b)$ Pulling out the time part out of the space part  \label{fig2}}
\vspace{-.2cm}
\end{figure}

 
To explain the partial derivative, Alex uses a ``pulling out'' motion, drawing it on the shared whiteboard (figure \ref{fig2}).  He explicitly talks about the assumption of separability and the role of partial derivatives:

\begin{description}

%

\vspace{-.3cm}

\item[Alex] We have $\frac{d}{dt}(f\psi)$, and then the $\psi$ comes out (while drawing a path showing the motion of coming out), so we just have $\psi$ partial derivative with respect to t 
\vspace{-.3cm}

\item[Eric] (surprised) Oh\dots We assume it's like this because (pointing to the $\Psi(x,t)$ = $\psi(x){f(t)}$)

\vspace{-.3cm}

\item[Alex] We assume it's separable. The assumption we are making is it [$\Psi$] can be separable and is function of $x$ times function of $t$

\vspace{-.2cm}

\end{description}


Alex on the bottom of the board writes the left side of the TDSE again to explain the procedure of pulling out.

Alex's use of word ``comes out'' indicates a mathematical metaphor as if the term is able to move and comes out of the parentheses. He also uses a hand gesture at the same time to display the specific path connecting the source (inside the parentheses) to the destination which is behind the derivative.

The use of gesture and metaphor in this procedural resource further indicates that Alex is trying to explain an action. Alex gesturally shows that some functions are constant such that the partial derivative does not act on them, and can be pulled out of the differentiation. 


At this point the assumption of the $\Psi(x,t)$ = $\psi(x){f(t)}$ makes sense to Eric as being consistent with the separated partial derivatives of space and time on each side of the TDSE.  Eric continues Alex's reasoning, remarking:

\begin{description}

\vspace{-.2cm}
\item[Eric] Which is kind of implied because you have this $t$ (pointing to the partial time derivative) on one side and this (pointing to the double partial space derivative) and there is nothing else.
\vspace{-.2cm}

\end{description}

Alex continues, writing the equation (see Fig.\ \ref{fig2}(b)) explicitly as $\frac{\partial (f\psi)}{\partial t} = -i\hbar \frac{\partial^2}{\partial x^2} f\psi$.  He  temporarily treats the $f$ as a constant to pull it out of the partial derivative on the right, saying ``then that comes out''. Eric gives voice to Alex's treatment of $f$ as a constant, confirming that   ``because that's [$t$ is] a constant, ok I got it\dots that makes sense.''  

We interpret this continued conversation as recruiting two more resources to unpack \textit{separation of variables}: the conceptual \textit{functions-as-constants} resource allows $f$ to act like a constant in light of the partial space derivative, and therefore Alex can use the procedural \textit{pulling out} resource to bring it outside the derivative.  From this point, Eric can continue with \textit{dividing} and \textit{separation constant} to complete the problem.  The fully unpacked resource graph for \textit{separation of variables} is in Figure \ref{fig1}c.




\section{Second interaction}

Interestingly, about two months later Alex and Eric become group mates again. In the first interaction, the potential was zero; in this interaction, the potential is a function of position and the problem is now in three dimensions, not one.  Otherwise the problem is the same.  Once again, the instructor wants the students to find the TISE in three dimensions for space part. 

Eric and Alex start the episode by discussing if the use of separation of variables is a good choice or not. They quickly run through the reasoning in Figure \ref{fig1}c, using the same hand gesture to show how functions can be treated as constants and pulled out of partial derivatives.  
Eric and Alex start the discussion with a valid assumption, using the idea that the potential and the Laplacian do not ``vary'' with time. Alex then uses a hand gesture showing a trajectory indicating the path of \textit{pulling out} resource and says ``so all carries out, it's separable''. At this point he does not mention, that what can carries out. Then he writes down the TDSE, \ecs{I don't know what this means} acting on the separated wave function $\psi(x)f(t)$, later he changes the label $x$ to $\textbf{r}$. He writes the time part on the left side of the equation, and the space part on the right side of the equation. After finishing writing the TDSE, Alex again uses a hand gesture, points to the $f(t)$  and says ``$f(t)$ carries out and then you can divide it over''.  Alex reuses the \textit{pulling out} resource to treat the function $f(t)$ as a constant that can carries out of the partial space derivatives. Alex similarly talks about the \textit{division} procedure after pulling out. He then points to the $\psi(x)$ on the left hand side of the equation and says ``and this is not a function of t, so you can divide it as well''.
  He divides both sides by $\Psi(x,t)$ and cancels the constant functions of each side of TDSE. Alex verbally reuses the \textit{pulling out} resource by showing hand gesture but, he does not draw a path on the white board. Alex thinks out aloud while solving the problem. 

However, in this episode Alex does not adopt the pedagogical tone he used in the prior interaction.  Eric does not ask for further clarification in the problem solving steps, and notices and corrects one of Alex's errors.  Their interaction seems more equitable.


We do not argue that the activation of \textit{pulling out} resource always goes along with gestural elements indicating the path of \textit{pulling out} resource. However, we observed that both Alex (and, in a separate interaction with another student, the instructor) used gestural elements to show how this resource works. We consider these gestural elements as evidence that pulling out is a procedural resource. 

\section{Discussion}

In this study, we observed that the right and left side of the time dependent Schr{\"o}dinger equation (TDSE) cues Alex to group one side as a function of time and the other side as a function of ${x}$, and to begin using \textit{separation of variables} to solve the problem.  In response to Eric's ongoing confusion about how to separate variables, Alex unpacks his procedural resources, iteratively adding more conceptual and procedural resources to make explicit the parts of \textit{separation of variables}.  Ultimately, the structure of Alex's \textit{separation of variables} includes three conceptual resources and three procedural.  


As we look across all of Alex's unpacking actions, we notice that Alex's explanation of the \textit{separation of variables} procedure is decidedly non-linear.  He starts with skipping the procedure altogether, preferring to assert a solution.  He expands to include two resources, then further details the link between them to describe two more.  When Eric is still confused, Alex eventually resorts to \textit{pulling out} and \textit{functions-as-constants} to unpack \textit{separation of variables} all the way.

More broadly, Thompson et al.~\cite{Thompson2006} investigated student difficulties with mixed partial derivatives. One common difficulty was application of the product rule in deriving the mixed second partial derivatives of thermal expansion, and thermal compressibility coefficients with respect to pressure, and temperature respectively. Students exhibit a variety of errors, factoring out some functions inappropriately and treating others as constants too often.  The simpler, unmixed partial derivatives in quantum mechanics better illustrate effective use of \textit{pulling out} and \textit{functions as constants}, and allow for better probes of the structure of students' \textit{separation of variables} resource, and thus how students can productively perform partial derivatives and solve PDEs.





In this study we identified a new procedural resource with a kinesthetic basis~\cite{Wittmann2013}. We investigated  students' use of ``pulling out'' resource in quantum mechanics  to convert the Schr{\"o}dinger partial differential equation to non-partial equations via separation of variables.    

The separated terms of partial  derivatives, with respect to time and space, in the TDSE cue students to group the equation based on two separated terms. However, not all of the students are aware that this is only correct if the wave function is separated into a function of only space and a function of only time. Using ``pulling out'' resource with gestural components is an efficient way for showing this elimination. We also showed that this resource is reusable as students after two months activate it again to solve a more general form of the TDSE with non-zero potential and in three dimensions successfully. The use of this resource is not limited only in partial differential equations in quantum mechanics; \textit{pulling out} is an integral part of solving PDEs across several upper-division courses. 

\acknowledgments{We thank participating instructors and students. This work is funded by NSF Grant DUE-1430967.}

\bibliographystyle{apsrev}  	


\begin{thebibliography}{9}
\expandafter\ifx\csname natexlab\endcsname\relax\def\natexlab#1{#1}\fi
\expandafter\ifx\csname bibnamefont\endcsname\relax
  \def\bibnamefont#1{#1}\fi
\expandafter\ifx\csname bibfnamefont\endcsname\relax
  \def\bibfnamefont#1{#1}\fi
\expandafter\ifx\csname citenamefont\endcsname\relax
  \def\citenamefont#1{#1}\fi
\expandafter\ifx\csname url\endcsname\relax
  \def\url#1{\texttt{#1}}\fi
\expandafter\ifx\csname urlprefix\endcsname\relax\def\urlprefix{URL }\fi
\providecommand{\bibinfo}[2]{#2}
\providecommand{\eprint}[2][]{\url{#2}}

\bibitem[{\citenamefont{Roundy et~al.}(2014)\citenamefont{Roundy, Kustusch, and
  Manogue}}]{Roundy2014}
\bibinfo{author}{\bibfnamefont{D.}~\bibnamefont{Roundy}},
  \bibinfo{author}{\bibfnamefont{M.~B.} \bibnamefont{Kustusch}},
  \bibnamefont{and} \bibinfo{author}{\bibfnamefont{C.}~\bibnamefont{Manogue}},
  \bibinfo{journal}{American Journal of Physics} \textbf{\bibinfo{volume}{82}},
  \bibinfo{pages}{39} (\bibinfo{year}{2014}).

\bibitem[{\citenamefont{Kustusch et~al.}(2014)\citenamefont{Kustusch, Roundy,
  Dray, and Manogue}}]{Kustusch2014}
\bibinfo{author}{\bibfnamefont{M.~B.} \bibnamefont{Kustusch}},
  \bibinfo{author}{\bibfnamefont{D.}~\bibnamefont{Roundy}},
  \bibinfo{author}{\bibfnamefont{T.}~\bibnamefont{Dray}}, \bibnamefont{and}
  \bibinfo{author}{\bibfnamefont{C.~A.} \bibnamefont{Manogue}},
  \bibinfo{journal}{Phys. Rev. ST Phys. Educ. Res.}
  \textbf{\bibinfo{volume}{10}}, \bibinfo{pages}{010101}
  (\bibinfo{year}{2014}).

\bibitem[{\citenamefont{Hammer}(|2000|)}]{Hammer2000}
\bibinfo{author}{\bibfnamefont{D.}~\bibnamefont{Hammer}},
  \bibinfo{journal}{American Journal of Physics} \textbf{\bibinfo{volume}{68}},
  \bibinfo{pages}{S52} (\bibinfo{year}{|2000|}).

\bibitem[{\citenamefont{Sayre and Wittmann}(2008)}]{Sayre2008}
\bibinfo{author}{\bibfnamefont{E.~C.} \bibnamefont{Sayre}} \bibnamefont{and}
  \bibinfo{author}{\bibfnamefont{M.~C.} \bibnamefont{Wittmann}},
  \bibinfo{journal}{Phys. Rev. ST Phys. Educ. Res.}
  \textbf{\bibinfo{volume}{4}}, \bibinfo{pages}{020105} (\bibinfo{year}{2008}).

\bibitem[{\citenamefont{Black and Wittmann}(2007)}]{Black2007}
\bibinfo{author}{\bibfnamefont{K.~E.} \bibnamefont{Black}} \bibnamefont{and}
  \bibinfo{author}{\bibfnamefont{M.~C.} \bibnamefont{Wittmann}},
  \bibinfo{journal}{AIP Conference Proceedings} \textbf{\bibinfo{volume}{951}},
  \bibinfo{pages}{53} (\bibinfo{year}{2007}).

\bibitem[{\citenamefont{Wittmann and Black}(2015)}]{Wittmann2015}
\bibinfo{author}{\bibfnamefont{M.~C.} \bibnamefont{Wittmann}} \bibnamefont{and}
  \bibinfo{author}{\bibfnamefont{K.~E.} \bibnamefont{Black}},
  \bibinfo{journal}{Phys. Rev. ST Phys. Educ. Res.}
  \textbf{\bibinfo{volume}{11}}, \bibinfo{pages}{020114}
  (\bibinfo{year}{2015}).

\bibitem[{\citenamefont{Griffiths}(2005)}]{griffiths2005}
\bibinfo{author}{\bibfnamefont{D.~J.} \bibnamefont{Griffiths}},
  \emph{\bibinfo{title}{Introduction to Quantum Mechanics}}
  (\bibinfo{publisher}{Pearson}, \bibinfo{year}{2005}).

\bibitem[{\citenamefont{Thompson et~al.}(2006)\citenamefont{Thompson, Bucy, and
  Mountcastle}}]{Thompson2006}
\bibinfo{author}{\bibfnamefont{J.~R.} \bibnamefont{Thompson}},
  \bibinfo{author}{\bibfnamefont{B.~R.} \bibnamefont{Bucy}}, \bibnamefont{and}
  \bibinfo{author}{\bibfnamefont{D.~B.} \bibnamefont{Mountcastle}},
  \bibinfo{journal}{2005 Physics Education Research Conference}
  \textbf{\bibinfo{volume}{818}}, \bibinfo{pages}{77} (\bibinfo{year}{2006}).

\bibitem[{\citenamefont{Wittmann et~al.}(2013)\citenamefont{Wittmann, Flood,
  and Black}}]{Wittmann2013}
\bibinfo{author}{\bibfnamefont{M.~C.} \bibnamefont{Wittmann}},
  \bibinfo{author}{\bibfnamefont{V.~J.} \bibnamefont{Flood}}, \bibnamefont{and}
  \bibinfo{author}{\bibfnamefont{K.~E.} \bibnamefont{Black}},
  \bibinfo{journal}{Educational Studies in Mathematics}
  \textbf{\bibinfo{volume}{82}}, \bibinfo{pages}{169} (\bibinfo{year}{2013}).

\end{thebibliography}

\end{document}